\def \yskip{\penalty-50\vskip3pt plus 3pt minus 2pt}
\def \reference{\par \yskip \noindent \hangindent .4in \hangafter 1}
\def \abc#1#2#3#4 {\reference#1, {\sl#2}, {\bf#3}, #4}
\def \blank {\lower 5pt\hbox to 0.75in{\hrulefill}}

\def \s{~\rm{s}}
\def \km{~\rm{km}}

\def \AU{~\rm{AU}}

\def \yrs{~\rm{yrs}}
\def \yr{~\rm{yr}}

\def \lesssim{\mathrel{<\kern-1.0em\lower0.9ex\hbox{$\sim$}}}
\def \gtrsim{\mathrel{>\kern-1.0em\lower0.9ex\hbox{$\sim$}}}

\documentstyle[11pt,aasms,tighten,flushrt]{article}

\begin{document}
\small

\setcounter{page}{1}

\begin{center}
\bf
COLLIMATED FAST WINDS IN WIDE BINARY \\
PROGENITORS OF PLANETARY NEBULAE
\end{center}

\begin{center}
Noam Soker\\
Department of Physics, University of Haifa at Oranim\\
Oranim, Tivon 36006, ISRAEL \\
soker@physics.technion.ac.il 
\end{center}


\begin{center}
\bf ABSTRACT
\end{center}

  We discuss the formation of planetary nebulae (PNe) having a
pair of lobes, or multi-lobes, in their inner region 
surrounded by an elliptical or spherical shell or halo.
Both elliptical and bipolar PNe are considered;
when the lobes are much smaller than the main elliptical shell,
the PN is termed elliptical PN, while when the lobes are the main
structure of the nebula, the PN is termed bipolar PN. 
 We suggest that most of these PNe are formed by wide
binary systems, with final orbital periods in the range of 
$\sim 40-10^4 \yrs$, such that there is no strong tidal interaction.
 The outer more spherical structure is formed from the early asymptotic
giant branch (AGB) wind.
 Toward the end of the AGB the mass loss rate increases, and wind velocity
possibly decreases, making the conditions for the formation of
an accretion disk around the wide companion more favorable.
 We assume that once a massive enough accretion disk is formed around 
the accreting companion it blows jets or a collimated fast wind (CFW),
which lead to the formation of a pair of lobes in the inner region.
 In cases of a precessing accretion disk, a multi-lobes structure
can be formed.
  We conduct a population synthesis study of such systems, and find
that overall $\sim 5-20 \%$ of all PNe are formed by such
binary systems.
 The exact percentage strongly depends on the wind velocities of
stars about to leave the AGB.  
 In about half of these systems the initially more massive star is
the AGB star and the accretor is a main-sequence star, while half have 
a WD accretor, with the initially less massive star being the AGB star.
 We also estimate that $\sim 20-40 \%$ of these systems 
possess observable departure from axisymmetry, e.g., the
central star is not in the center of the nebula.
  Our population synthesis not only supports the binary model for
formation of these types of PNe, e.g., Hu2-1, He 2-113, He 2-47,
and M1-37, but more generally supports the binary model for the
formation of bipolar and many elliptical PNe.

{\it Subject headings:}
planetary nebulae: general
$-$ stars: binaries
$-$ stars: AGB and post-AGB
$-$ stars: mass loss
$-$ ISM: general


\section{INTRODUCTION}

 New ground and space telescopes make possible high resolution
observations of planetary nebulae (PNe; e.g., Sahai \& Trauger 1998,
Balick {\it et al.} 1998, Goncalves, Corradi \& Mampaso 2001).
 These observations resolved previously known features into fine details
and revealed new features, imposing strong constraints on models
for the formation of aspherical PNe.
 Some of the features which turned out to be present in many PNe are
narrow lobes, jet-like features, and dense blobs, which in many cases
are located within an almost spherical halo (e.g. NGC 6826,
Plait \& Soker 1990; The Egg Nebula, Sahai {\it et al.} 1998;
IRAS 04296+3429, Sahai 1999; NGC 6572, Miranda {\it et al.} 1999;
NGC 6891, Guerrero {\it et al.} 2000; M1-16, Huggins {\it et al.} 2000;
NGC 6543, Balick, Wilson, \& Hajian 2001; more images in Sahai 2001).
 In many cases there are more than one pair of lobes, or jet-like
features, pointing in different directions which hint to
precessing jets or collimated flow (e.g., Sahai 2000, 
Miranda {\it et al.} 2001a).
 These multipolar structures can't be explained by the interaction of a
spherical fast wind, blown by the central star, with the previously
ejected slow wind (e.g., Soker 1990; Sahai, Nyman, \& Wootten 2000);
rather, the jets that formed the multipolar structures and dense blobs
(sometimes called {\it ansae}) are formed close to one of the stars
in a binary system, and in many systems are blown during the transition
from the AGB to the PN phase (Soker 1990; Sahai \& Trauger 1998).
 The details seen in many PNe and proto-PNe made some people 
claim that current models can't explain some of these structures.
 An example is the claim made by Balick {\it et al.} (1998) that
fast bullets can't explain some properties of FLIERs
(Fast Low-Ionization Emission Regions), a claim that is disputed
by Soker \& Regev (1998).
 Another example is the claim made by Sahai {\it et al.} (2000), who
argue that the offset of the central star of the PN He 2-113 from
the center of the nebula is puzzling.
  In the present paper we argue that wide binary systems can naturally
form the different structures observed in the PNe having multipolar
structures within a spherical, or almost spherical, halo.
 In particular, in  many systems a companion to the progenitor of the
elliptical PN blows jets or a collimated fast wind (CFW), which
can form the lobes and ansae.
 The symbiotic star HM Sge show that binary systems with no strong
tidal interaction can still lead to the formation of aspherical nebulae.
 This symbiotic system has an orbital separation of $\sim 50 \AU$
(Eyres {\it et al.} 2001), and possess a very elongated, possibly bipolar,
nebula around it.
 We also show here that many elliptical PNe are likely to possess departure
from axisymmetry, a claim that was raised in some of our
earlier papers (Soker \& Rappaport 2001 and references therein).
 The wide binary systems, though, don't form all elliptical PNe, but
only a fraction of them, as we explain in section 4. 
 The different parameters that characterized binary systems,
e.g., orbital semimajor axis, eccentricity, masses of the
two stars, composition, presence of planets close to one of the stars,
ensure a rich variety of different PNe structures.
 To find this rich spectrum of shapes one needs to explore the
different regions of the binary system parameters space.

 Another common property of many PNe and proto-PNe, as is seen in many of
the images cited above (e.g., Sahai 2001), is that the transition from
a spherical (or almost spherical) to an axisymmetric mass loss occurs
at the same time as the mass loss rate increases substantially.
 The high mass loss rate phase is sometimes referred to as the
superwind phase.
 However, the term superwind is also used to describe equatorially
enhanced winds.
 To avoid confusion, we will term the high mass loss rate which produces
an optically thick wind `FIW', for `final intensive wind'.
 PNe which show no signatures of FIW tend to be spherical (Soker 1997).
 The correlation between the transition to axisymmetric mass loss and
FIW was attributed to properties of a single star, which probably have
been spun-up by a stellar or substellar companion, or to a late 
interaction with a companion (see review by Soker 2000a). 

  In the present paper we examine the possibility that in a fraction
of progenitors of elliptical PNe the FIW may initiate the formation of
an accretion disk around a wide companion, leading to the formation of
a CFW (or jets) blown by the companion
(Soker \& Rappaport 2000, hereafter SR00).
 This CFW then shapes the inner region of the circumstellar material,
and later manifests itself in the shape of the proto-PN and the
PN itself.
  We must define first the terms bipolar and elliptical PNe, which
will be used throughout this paper.
 We adopt the definition of bipolar (also called ``bilobal'' and
``butterfly'') PNe as proposed by Schwarz, Corradi \& Stanghellini (1992),
as axially symmetric PNe having two lobes with an ``equatorial''
waist between them.
 Elliptical PNe have a general elliptical shape, with no waists, but
may possess lobes and jets within, or outside, the main elliptical
nebula.
 According to this definition He 2-113 is an elliptical PN,
and not a bipolar PN as called by Sahai {\it et al.} (2000).

  In section 2 we discuss wide binary systems, in which the
accretion process by the companion may form an accretion disk
and blow a CFW.
 In section 2 we show also that the proposed model for transition
from spherical (or almost spherical) to axisymmetric mass loss,
in many systems naturally leads to a departure from axisymmetry
of the descendant PN, as is the case for many elliptical PNe.
  In section 3 we conduct a population synthesis to estimate the number
of wide-binary progenitors of elliptical PNe which are expected to
change their mass loss from spherical to axisymmetrical during
the FIW phase, via the formation of a CFW.
  A large fraction of this paper is an extension of SR00.
 For clarity in sections 2 and 3 we present some, but not all,
of the assumptions and definitions from SR00.
 More detail can be found in SR00 and Soker \& Rappaport (2001).
 We discuss and summarize our main results in section 4.

\section{THE WIDE BINARY PROGENITOR MODEL}

  It is generally agreed that a disk is a necessary condition for the
formation of jets emanating from compact objects
(for a recent review see Livio 2000).
 In the model proposed by SR00, the CFW can have a wide
opening angle, and is not limited to a narrow jet, hence
the properties of the disk are not limited, e.g., that it be extended,
but rather it is only required that an accretion disk be formed
around the accreting star.
  This condition reads $j_a > j_2$, where $j_a$ is the specific
angular momentum of the accreted material, and $j_2=(G M_2 R_2)^{1/2}$
is the specific angular momentum of a particle in a Keplerian orbit at
the equator of the accreting star of radius $R_2$ and mass $M_2$.
 The expressions we use below hold both for the case when the accretor
is the WD remnant of the initial primary star or the accretor is the
main-sequence secondary.
 The mass and radius are dramatically different in these two cases, though.
 For accretion from a wind, the net specific angular momentum of the
material entering the Bondi-Hoyle accretion cylinder with radius $R_a$,
i.e., the material having impact parameter
$b<R_a = 2 G M_2 /v_r^2$, where $v_r$ is the
relative velocity of the wind and the accretor, is (Wang 1981)
$j_{BH} = 0.5 (2 \pi / P_o) R_a^2$, where $P_o$ is the orbital period.
  Livio {\it et al.} (1986; see also Ruffert 1999) find that the actual
accreted specific angular momentum for high Mach number flows is
$j_a = \eta j_{BH}$, where $\eta \sim 0.1$ and $\eta \sim 0.3$
for isothermal and adiabatic flows, respectively.
 The relative velocity is $v_r^2 \simeq v_s^2 + v_o^2$, where $v_s$
is the (slow) wind velocity at the location of the accreting star, and $v_o$
is the relative orbital velocities of the two stars.
 Substituting typical values for WD accretor and the mass-losing
terminal AGB star we find the following condition for the
formation of a disk
\begin{eqnarray}
1< \frac {j_a}{j_2} = 15
\left( \frac {\eta}{0.2} \right)
\left( \frac {M_1+M_2}{1.2 M_\odot} \right)^{1/2}
\left( \frac {M_2}{0.6 M_\odot} \right)^{3/2}
\left( \frac {R_2}{0.01 R_\odot} \right)^{-1/2}
\left( \frac {a}{10 \AU} \right)^{-3/2}
\left( \frac {v_r}{15 \km \s^{-1}} \right)^{-4} ,
\end{eqnarray}
where the expression is for a circular orbit with a semimajor axis $a$.
(Note that the exponent $1/2$ is missing in the second parentheses
of equation 2 in SR00.)
 Another plausible condition for the formation of a CFW is that the
accretion rate should be above a certain limit $\dot M_{\rm crit}$,
which we take as $10^{-7} M_\odot \yr^{-1}$ for accretion onto a
main-sequence star and $10^{-8} M_\odot \yr^{-1}$ for accretion
onto a WD (SR00).
 The accretion rate depends on several factors
(Mastrodemos \& Morris 1999), e.g.,
the accretor mass, the acceleration zone of the slow wind,
concentration toward the equatorial plan, and synchronization of
the mass-losing star.
 We neglect most of these effects, and take the accretion
rate to be $\dot M_2 = \pi R_a ^2 v_r \rho$,
where the density at the location of the accretor is
$\rho = \vert \dot M_1 \vert / (4 \pi a^2 v_s)$.
 Substituting the relevant parameters during the FIW phase, i.e.,
the high mass loss rate phase at the end of the AGB, we get
\begin{eqnarray}
\dot M_2 \simeq
5.5 \times 10^{-6}
\left( \frac {M_2}{0.6 M_\odot} \right)^{2}
\left( \frac {v_r}{15 \km \s^{-1}} \right)^{-3}
\left( \frac {v_s}{15 \km \s^{-1}} \right)^{-1}
\left( \frac {a}{10 \AU} \right)^{-2}
\left( \frac {\vert \dot M_1 \vert }{10^{-4} M_\odot \yr^{-1}} \right)
M_\odot \yr^{-1}.
\end{eqnarray}

 Let us analyze the angular momentum and accretion rate conditions
for the formation of an accretion disk and a CFW or jets.
 Both expressions depend strongly on the orbital separation, and very
strongly on the wind velocity $v_s$, which appears in $v_r^2=v_s^2+v_o^2$.
 The accretion rate condition also depends on the mass loss rate.
 These dependencies lead to the following behavior of the CFW in binary
systems where  $a \gtrsim 10 \AU$ (much closer binary systems will go
through Roche lobe overflow and/or common envelope phases; SR00). 
\newline 
{\bf 1. Aspherical inner region inside a spherical halo.}
A CFW is formed only when the mass-losing star reaches the
upper AGB, when mass loss rate is high and the wind is slow
(in some cases a CFW might be formed during the first red giant branch
[RGB]).
 Because of the very strong dependence on the wind velocity, in
many binary systems the accretor blows a CFW only when the
wind velocity decreases substantially.
 A decrease in the wind velocity may occur during the FIW phase,
when the wind is optically thick, hence the AGB stellar radiation
accelerates the wind along a short distance above the stellar surface.
 If this is the case, in many systems a CFW will be formed
only at the very final stages of the AGB (the FIW phase).
 Even if a CFW is presence before the FIW phase, because of the
decrease in the slow wind velocity and the increase in the
accretion rate, the ratio of the momentum flux of the CFW to
the momentum flux of the slow wind will increase.
 This will increase the role of the CFW in shaping the descendant PN
(SR00).
 In all these cases, the descendant PNe will have a very non-spherical
inner structure embedded inside a larger more spherical nebula.
 The inner highly non-spherical region will be much denser with respect
to the halo than in the $r^{-2}$ constant wind density case, since
it is formed from a more intense wind (the FIW).

\noindent {\bf 2. Departure from axisymmetry.}
 The strong dependence on the orbital separation of both the mass
accretion rate and angular momentum conditions means that in many systems
an accretion disk, hence a CFW (or jets) by our assumption, is formed
near periastron passages but not near apastron passages, i.e.,
only along a fraction of the orbital motion.
 This possibility has been suggested to occur in Hu2-1
(Miranda {\it et al.} 2001b, their $\S 3.5$).
 Even if a CFW is blown during the entire orbital period, it is
expected to be stronger near periastron passages. 
 This kind of non-axisymmetrical behavior leads to a departure
from axisymmetry of the descendant PN (Soker \& Rappaport 2001).
 Even if the CFW strength is constant, in systems where the orbital period
is of the order of the duration of the FIW a departure from axisymmetry
is expected in the descendant PN (Soker 1994; Soker \& Rappaport 2001). 
 In recent papers Miranda {\it et al.} (2001b; for Hu2-1)
and Miranda, Guerrero, \& Torrelles (2001a; for IC 4846), proposed that the
difference between the systematic velocity of the precessing jets and the
centroid velocity of the nebulae in these two PNe results from the
orbital motion of the star that blows the jets.
 This is one of the manifestations of departure from axisymmetry
in binary systems (Soker \& Rappaport 2001).

\noindent {\bf 3. Delayed CFW.}
 When the slow wind in the equatorial plane is very slow,
$v_s \simeq 5 \km \s^{-1}$, in a small number of systems,
with a WD or a main-sequence companion, a CFW is formed at
orbital separations as large as $\sim 500 \AU$ (see next section).
 The time required for such a wind to flow from the AGB mass-losing star
to the accretor is $\sim 500 \yrs$.
 This means that the CFW will be operating for hundreds of years
after the mass-losing star has left the AGB.

\noindent{\bf 4. Elliptical PNe with prominent equatorial structure.}
 This class of PNe is defined by SR00 (their section 3.3), as PNe
where a CFW is formed, but its momentum density flux is small
so that it is sharply bent toward the equatorial plane by the
slow wind.
 In such a case the CFW shapes only regions close to the equatorial
 plane.
If the slow wind's velocity is low, and the CFW is well collimated,
i.e., two jets, the CFW isn't bent much, and influences the
descendant PN structure near the polar directions.
 The population synthesis can't teach us when the CFW is sharply bent
and when it expands along the polar directions.
 In any case, if the CFW is bent, the structure may possess a
mirror symmetry around the polar direction.
 This can be in addition to a point-symmetric structure which may result
from precessing jets (Livio 2000).

 In section 3 we examine the situation when the wind velocity decreases
during the FIW phase.
 We now try to justify this assumption, namely that during an optically
thick wind there is a {\it negative} correlation between the mass loss
rate and wind velocity.
 Because of the optically thick wind, there are no observation of
the wind during the FIW phase.
 For optically {\it thin} winds, there is a {\it positive} correlation
between the mass loss rate and the wind's velocity
(Sahai \& Liechti 1995).
 This is explained by the presence of dust; the more efficient the
dust formation process is, the more efficient is the acceleration
by radiation pressure, which increases both mass loss rate
and wind velocity (Sahai \& Liechti 1995).
This is the case in S stars, where both wind velocity and mass loss rate
are low.
 For our purpose, the wind velocity in S stars, which can get as low as
$\sim 5 \km \s^{-1}$ (Sahai \& Liechti 1995), demonstrates that winds
from AGB stars can be much slower than the escape velocity.
 Another indication is the very slow expansion velocity,
$\lesssim 10 \km \s^{-1}$, observed in some young PNe,
e.g., the extremely young PN Vy2-2, where the expansion velocity,
probably in the equatorial plane, is $6 \km \s^{-1}$
(Miranda \& Solf 1991).
 Another possible indication for a negative correlation between mass
loss rate and velocity during the FIW is the structure of some
elliptical PNe and proto-PNe, where the dense equatorial flow
expands much slower than the rarefied polar flow,
even before the fast wind interacts with the slow wind.
 We therefore consider very plausible our assumption that the wind
velocity decreases during the FIW, when the wind is optically thick
close to the star.
 This may be particularly the case above cool spots where large
quantities of dust are formed (Soker 2000b).
 In any case, our main conclusions hold for the case when the
wind velocity does not change along the upper AGB, although they
are less pronounced.

\section{POPULATION SYNTHESIS}
\subsection {Specific Prescription and Algorithms}

 We use the same binary systems that were simulated by SR00,
where a description of the Monte Carlo type population synthesis
code is given.
  We examine the relevant systems at their final evolutionary point,
when the mass-losing star has lost almost its entire envelope, since
we are interested in the FIW phase.
 There are four main differences between the present study and
the results presented by SR00.
\newline
(1) SR00 followed the entire evolution of the mass-losing star.
 Here we examine the relevant systems taken from SR00 at their
final evolutionary point, when the mass-losing star retains a
very low mass envelope, i.e., the mass of the mass-losing
star is its core mass. 
 We do this since our interest is the FIW phase at the termination
of the AGB.
\newline
(2) SR00 were interested in the formation of bipolar PNe.
Therefore, they were mainly interested in systems where the companion
star is outside the AGB envelope and the systems have a strong tidal
interaction, i.e., the binary system reaches tidal synchronization and
circularization.
 These are class C (when the initial more massive star is the AGB star;
hereafter termed primary-AGB cases)
and class D (where the AGB star is the initially less massive star;
hereafter termed secondary-AGB cases)
in Table 2 of SR00.
We are interested mainly in elliptical PNe, so we present results mainly
for systems that do not reach synchronization and circularization.
 These are classes A (primary-AGB cases) and B (secondary-AGB cases)
in Table 2 of SR00
\newline
(3) To avoid complications, and since they were interested mainly
in tidally strongly interacting systems, in the criteria for the
formation of a CFW (eqs. 1 and 2 above) SR00 took the relative
velocity between the accreting star and the wind to be
$15 \km \s^{-1}$, and neglected the difference between the wind
velocity $v_s$ and relative velocity $v_r$.
 Here we take a more accurate expression for the relative velocity 
$v_r = (v_s^2 + v_o^2)^{1/2}$, where $v_o$ is the relative orbital
velocity of the two stars.
 We also examine the role of the wind velocity.
\newline
(4) For the same reason as above, $a$ in the criteria for the
formation of a CFW SR00 took to be the semimajor axis $a_0$.
 Here we examine separately the criteria at periastron, where
the distance in the criteria for the formation of a CFW is
$a=a_0(1-e)$, and at apastron, where $a=a_0(1+e)$. 

 Our goal is to find the number of PNe, which will be
presented as the percentage of the total number of PNe, which
blow CFW as a function of the wind velocity and mass loss rate.
 We shall also present the distribution of orbital periods
for systems which do not have strong tidal interaction
(do not reach circularization and synchronization),
but blow a CFW, either during a fraction of the
orbital motion (near periastron) or during the entire orbit.
 Most of them are likely to form elliptical PNe, but some
form bipolar PNe (SR00 Table 2).
 If the CFW becomes strong at the final stage of the AGB,
a very aspherical structure will be present in the inner
regions of the descendant proto-PNe and PNe.
  Some binary systems will form two PNe; one is formed by the initial
more massive star, and much later a PNe is formed by the initially
less massive star.
 Some single stars will also form PNe as well.
 On the other hand many single and binary systems with low mass stars
will not form any PN, so we assume that on average for each
binary system one PN is formed (see section 4.3 of SR00 for
detailed discussion).
 The total number of binary systems at the beginning of the
population synthesis run is 46,700 (SR00 section 5.1).

\subsection{Population Synthesis Results}

 As is expected from equation 2, the number of systems blowing CFW
increases as the mass loss rate increases, because the
condition for blowing a CFW is met at larger distances.
 However, this is true up to a limit mass loss rate, at which point
the condition on the angular momentum, equation 1, limits the
formation of a CFW.
 The number of systems which blow a CFW increases above this mass
loss rate only when the wind velocity decreases.
 This expectation is presented quantitatively in figures 1-5,
which present the variation of the number of systems that
blow CFW, expressed as a percentage of the total number of binary
systems we start with, as a function of the mass loss rate by the
AGB star, or the wind velocity.
We assume that this percentage is about equal to the percentage out of
all PNe (SR00).
 In Figures 1-3 and 6-8 the sample is divided according to whether 
the initial more massive star (primary AGB) or the initial less
massive star (secondary AGB) is the AGB star.
 Thin lines (marked ``Periastron'') represent systems where a CFW is
blown only during a fraction of the orbit, when the orbital separation is
smaller, i.e., around periastron,
while thick lines (marked ``Orbit'') depict systems which blow CFW
along the entire orbit. 

 Although we are mostly interested in progenitors of elliptical PNe,
in Figure 1 we present the results for progenitors of bipolar
PNe (classes C and D of SR00).
 These are systems that in addition to blowing a CFW have strong
tidal interaction, so that they reach circularization and synchronization,
but they do not enter a common envelope phase or a Roche lobe overflow. 
 For these the number of systems does not increase with
increasing mass loss rate beyond $\sim 10^{-5} M_\odot \yr^{-1}$
for two reasons.
 The first is the criterion on angular momentum, as mentioned above.
 The second is the requirement that these systems have strong tidal
interaction, hence these systems are limited to relatively small
orbital separation. 
 Most systems that blow CFW do so already at mass loss rates
of $\lesssim 10^{-6} M_\odot \yr^{-1}$, much before the
mass-losing star reaches the FIW phase, hence forming bipolar PNe.
  These systems are more complicated since the AGB star rotates
at the system angular velocity, and may blow a stronger wind
in the equatorial plane.

 Figures 2-8 present results for systems that blow CFW, but do not
have strong tidal interaction, hence do not reach circularization
and synchronization. These are termed class A (for primary AGB)
and B (for secondary AGB) by SR00.
 Most of these systems form elliptical PNe, and a minority are
likely to form bipolar PNe (SR00).
 We note that the number of these systems that blow CFW steeply
increases as the star enters the FIW phase, when mass loss rate
increases to $\gtrsim 3\times 10^{-6} M_\odot \yr^{-1}$.
 This will be even more pronounced if the wind velocity decreases
during the FIW phase, as we propose may happen (see previous section).
 The descendant PNe will have an almost spherical halo, with a
strongly non-spherical inner and brighter region. 

 In figures 6-8 we present the distribution (number of systems in the
simulation rather than percentage) of the orbital periods for systems
with no strong tidal interaction and for three different wind
velocities.
 We see that the formation of a disk around a WD is more favorable
than around a main-sequence star, as is expected from equation (1).
 We also see that many CFW systems have periods of the same order
as the duration of the FIW phase, $\sim few\times 1,000 \yrs$.
 In these systems a significant departure from axisymmetry is
expected (Soker 1994; Soker {\it et al.} 2001).
 
\section{DISCUSSION AND SUMMARY}

 Before analyzing the results we stress the implication
of our assumption of a spherical mass loss by the mass-losing AGB star.
 If the mass loss rate is higher in the equatorial plane,
due to the rotation of the AGB star for example, the mass flowing
toward the companion is higher than under the assumption of a spherical
mass loss.
 This will increase both the likelihood of forming two jets or a
collimated fast wind (CFW), and the strength of the jets, or CFW,
relative to the slow wind blown by the AGB star perpendicular to
the equatorial plane.
 The implications of this uncertainty in the exact mass-loss geometry,
however, are smaller than the implications of the uncertainties in the
exact values of the wind velocity, and the conditions for the
formation of an accretion disk and jets or CFW. 

 Taking these uncertainties into consideration, we arrive at the following
conclusions.
 Although many systems blow CFW or jets only during a fraction of
the orbit (marked ``Periastron'' in the figures), for wind velocities
$v_s \gtrsim 10 \km \s^{-1}$ the orbital periods (Figures 6-7) are
such that most of them will reach periastron during the final intensive
wind (FIW) phase.
 For $v_s \sim 5 \km \s^{-1}$ the orbital periods in some systems may be
much longer than the FIW phase of few$\times 1,000 \yrs$, and these
systems may be close to apastron during the FIW phase, hence they
will not blow jets of CFW.
 However, as can be seen from Figure 8, there are not many such systems.
 We note also that according to SR00 $\sim 0.5 \%$ (of all PNe)
of this type of systems form bipolar PNe rather than elliptical PNe
(see their Table 2). 
 Considering these factors, and based on Figures 4 and 5 we estimate
that $\sim 5-20 \%$ of all PNe belong to the type of systems we study
here, namely systems which blow CFW or jets only at the final stage
of the AGB, and have no strong tidal interaction.
 The exact percentage strongly depends on the wind velocities of
stars about to leave the AGB.  
 About half of the systems will be primary-AGB cases, i.e., the
initially more massive star is the AGB and the accretor is a
main-sequence star, while half will have a WD accretor, with the
initially less massive star being the AGB star.
  Because of the weak tidal interaction and the fact that the jets or CFW
are blown only at the termination of the AGB, most of the mass of
the descendant PNe will be elliptical or spherical, but with lobes
and/or jets in the inner region (at later stages of the PN phase
the CFW or jets will break outward, though).
 We suggest this model to explain PNe and proto-PNe such as
He 2-113 (Sahai {\it et al.} 2000; more examples are in the first
paragraph of section 1 above).
 We note that the precession of the accretion disk can lead to the
formation of multipolar structure (Livio 2000). 
 
 As evident from Figures 6-8, many systems, the number increases with
deceasing wind velocity, have long orbital periods, such that the
systems will complete at most a few orbital periods during the FIW.
 The descendant PNe of these systems are expected to possess a
significant departure from axisymmetry (Soker 1994).
 The departure from axisymmetry will be more prominent in systems having
eccentric orbits where a CFW (or jets) is blown only at a fraction of the
orbit (Soker {\it et al.} 2001).
  A very good example is Hu2-1 (Miranda {\it et al.} 2001b),
where observations suggest an eccentric orbit and a CFW which is
stronger near periastron passages ($\S 3.5$ of Miranda {\it et al.}
2001b).
 Based on Figures 6-8, assuming $\sim 3,000 \yrs$ for
the average duration of the FIW and the results of Soker {\it et al.}
(2001), we crudely estimate that $\sim 20-40 \%$ of the systems studied
here will possess observable departure from axisymmetry.
 A departure from axisymmetry, in that the central star is not at
the center of the nebula, is noted by Sahai (2000) in the two PNe
He 2-47 and M1-37, which also show multipolar structure inside
a more spherical halo. These 2 PNe are example of the type
of PNe we expect to be descendant of the binary systems studied in
the present paper. 

 We can put the estimates of the present paper in the context of
all PNe.
  With $\sim 10 \%$ of all PNe being spherical (Soker 1997),
and $10-15 \%$ bipolar (Corradi \& Schwarz 1995), the fraction of
elliptical PNe is $\sim 75-80 \%$.
 In the present paper we find that $\sim 5-20 \%$ of all PNe,
or $\sim 6-25 \%$ of all elliptical PNe are likely to have a sharp
transition from an outer spherical and faint halo to an inner highly
aspherical and bright region, as a result of wide binary interaction.
 A large fraction of these will possess observable departure from
axisymmetry.
 Note that other processes can lead to a nebular structure of a
highly non-spherical and dense region within a larger faint and
more spherical structure, e.g., a steep increase in the efficiency
of cool magnetic spots formation (Soker 2000a).
 These processes, though, are less likely to form lobes, especially
multipolar lobes.

 The possibility of a very slow wind at the final AGB phase
adds $\sim 15 \%$ to the number of PNe that acquire their
axisymmetrical structure from a binary companion;
here the companion accretes and blows a CFW or jets. 
 We find the number of systems that blow CFW (or jets)
to be $3\%$ for $v_s=15 \km \s^{-1}$, similar to the $3.5 \%$
found by SR00, their classes A and B (their Table 2), while we find 
$12.5\%$ for $v_s=10 \km \s^{-1}$, and
$23.5\%$ for $v_s=5 \km \s^{-1}$.
 The extra $\sim 15 \%$ binary systems reduce the number of
progenitors which need to acquire their axisymmetrical
structure from planets or fastly rotating single stars
from $\sim 55 \%$ (Soker 1997) to $\sim 40 \%$.
 
\bigskip

{\bf ACKNOWLEDGMENTS:}
I thank Luis Miranda for very helpful comments.
 This research was supported in part by grants from the
US-Israel Binational Science Foundation.



\newpage

{\bf FIGURE CAPTIONS}

\noindent {\bf Figure 1:}
 The fraction of binary systems for which the companion to the AGB
stars blows a collimated fast wind (CFW) vs. the mass loss rate by
the AGB star. The conditions for the formation of a CFW are given by
equations 1 and 2. The fraction is given as percentage of
the total number of binary systems at the start 
(46,700 systems; see SR00 for details).
 The results are only for systems with a strong tidal interaction
that brings the systems to circularization and synchronization,
but the systems avoid common envelope and Roche lobe overflow.
 Although the systems reach circularization, for comparison with
Figures 2-3 for systems with weak tidal interaction, we did not
change the eccentricity of these systems when presenting the
results here.
 The top panel refers to such systems that formed when the original
primary star in the binary ascends the AGB and the companion is
the original secondary (termed class C by SR00),
while the lower panel is the same for the case where the secondary is
an AGB star and the companion is the white dwarf remnant of the
original primary (termed class D by SR00).
 These systems are the principal candidates for forming very narrow
waist bipolar PNe in binary systems (SR00).
 The thick lines are for systems where a CFW is blown during the entire
orbital period, while the thin line depicts systems where a CFW is
blown during a fraction of the orbit only, i.e., no CFW is
blown near apastron.
 The wind velocity is $10 \km \s^{-1}$. 
 The wiggling of the lines shows the statistical noise of the
numerical code. 

\noindent {\bf Figure 2:}
 Same as Figure 1, but for systems which have weak tidal interaction
so don't reach circularization
(these are termed class A and B by SR00), and for $v_s=15 \km \s^{-1}$.
Most of these systems will form elliptical PNe, while the minority
will form bipolar PNe (see Table 2 by SR00 for details).

\noindent {\bf Figure 3:}
 Same as Figure 2, but for a wind velocity of $v_s=10 \km \s^{-1}$.

\noindent {\bf Figure 4:}
 The total fraction of systems having weak tidal interaction and
blowing a CFW vs. the mass loss rate of the AGB star.
 The CFW can be blown during the entire orbit or a fraction of it,
for a wind velocity of $v_s=10 \km \s^{-1}$ (upper line)
and a wind velocity of $v_s=15 \km \s^{-1}$ (lower line).
 The upper line is the sum of the 4 lines in Figure 3 and
the lower line is the sum of the 4 lines of Figure 2.

\noindent {\bf Figure 5:}
 Same as Figure 4, but vs. the wind velocity and for two values of
the AGB star mass loss rate, $\vert \dot M_1 \vert = 10^{-6}$ and 
$10^{-5} M_\odot \yr^{-1}$, as indicated.

\noindent {\bf Figure 6:}
Distributions of final orbital periods for systems which blow
CFW when the AGB mass loss rate and wind velocity are
$\vert \dot M_1 \vert = 10^{-5} M_\odot \yr^{-1}$ and
$v_s=15 \km \s^{-1}$, respectively.
 The total number of binary systems at the start is 46,700.
  The thick lines are for systems where a CFW is blown during the entire
orbital period, while the thin line depicts systems where a CFW is
blown during a fraction of the orbit only, i.e., no CFW is
blown near apastron.
 The top panel refers to such systems that formed when the original
primary star in the binary ascends the AGB and the companion is
the original secondary, while the lower panel is the same for the
case where the secondary is an AGB star and the companion is the
white dwarf remnant of the original primary.

\noindent {\bf Figure 7:}
 Same as Figure 6, but for a wind velocity of $v_s=10 \km \s^{-1}$.

\noindent {\bf Figure 8:}
 Same as Figure 6, but for a wind velocity of $v_s=5 \km \s^{-1}$.

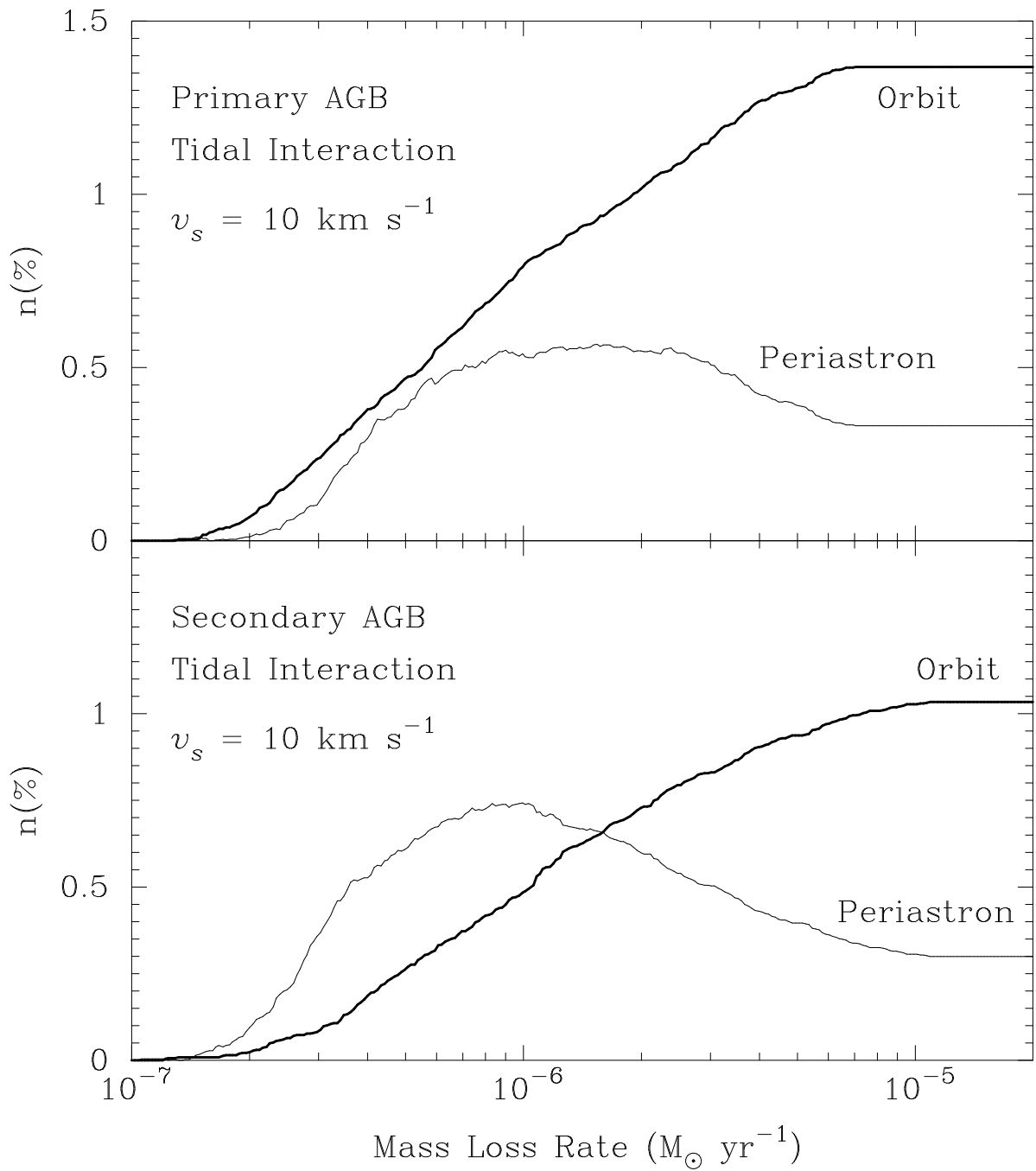
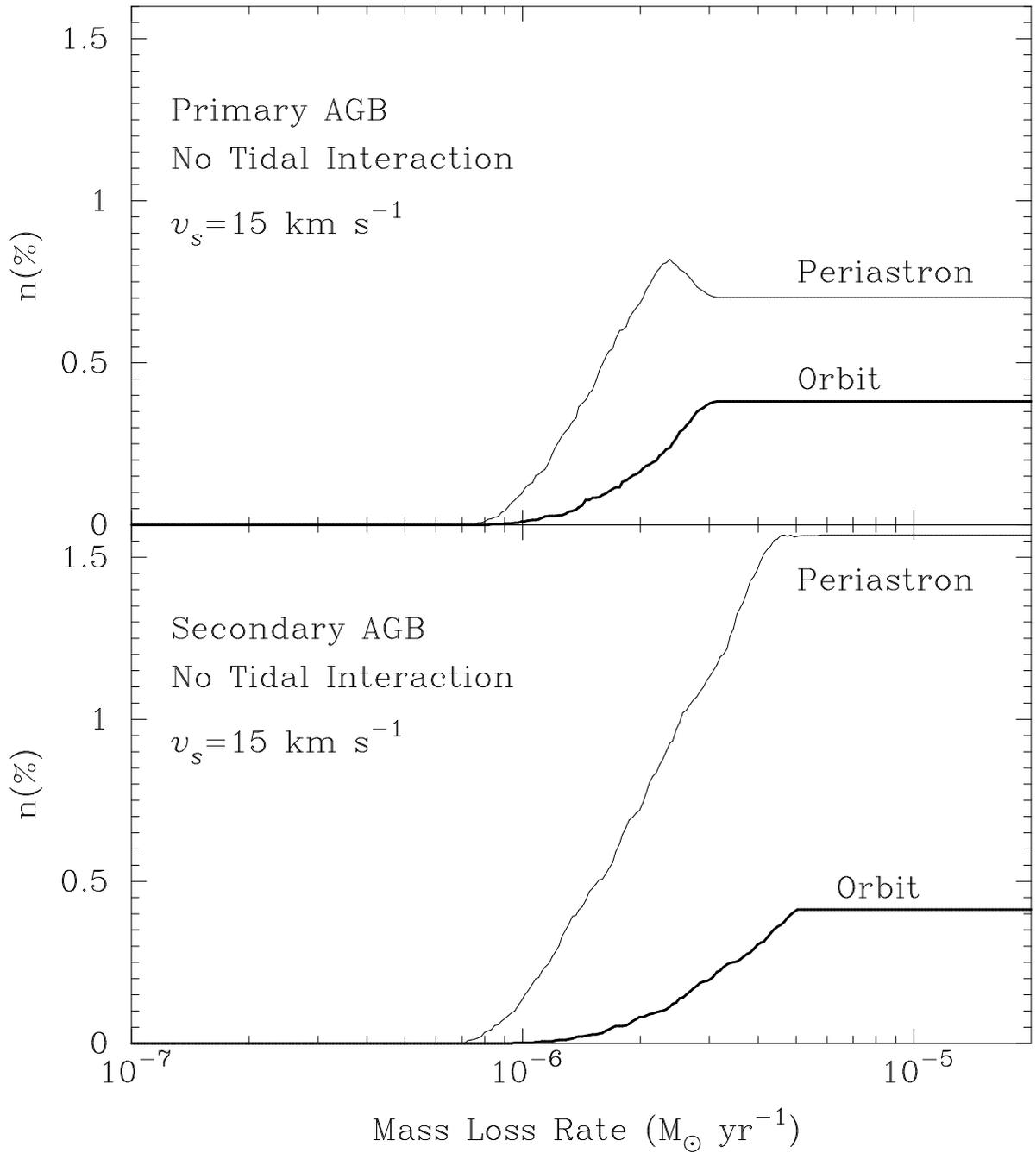
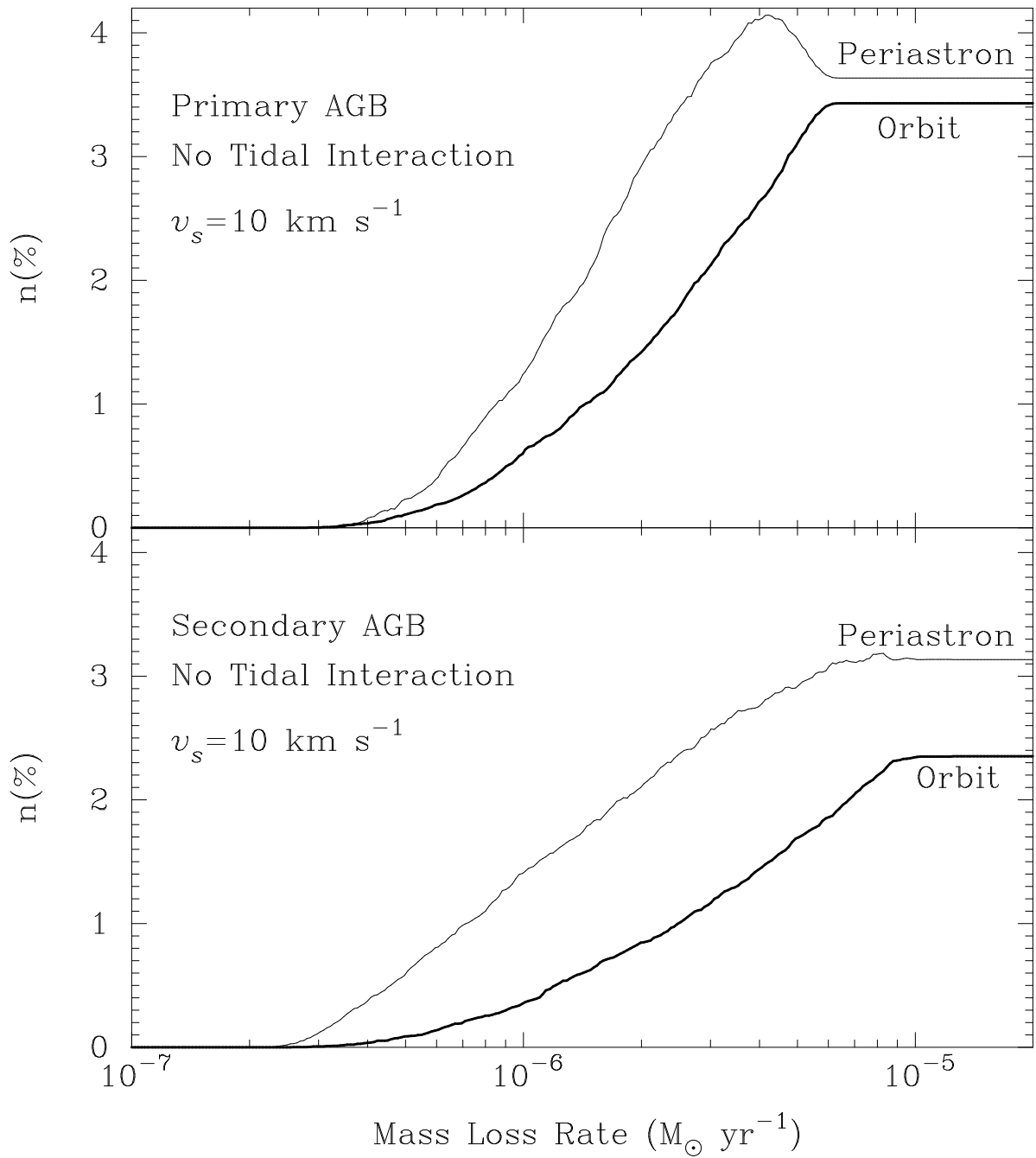
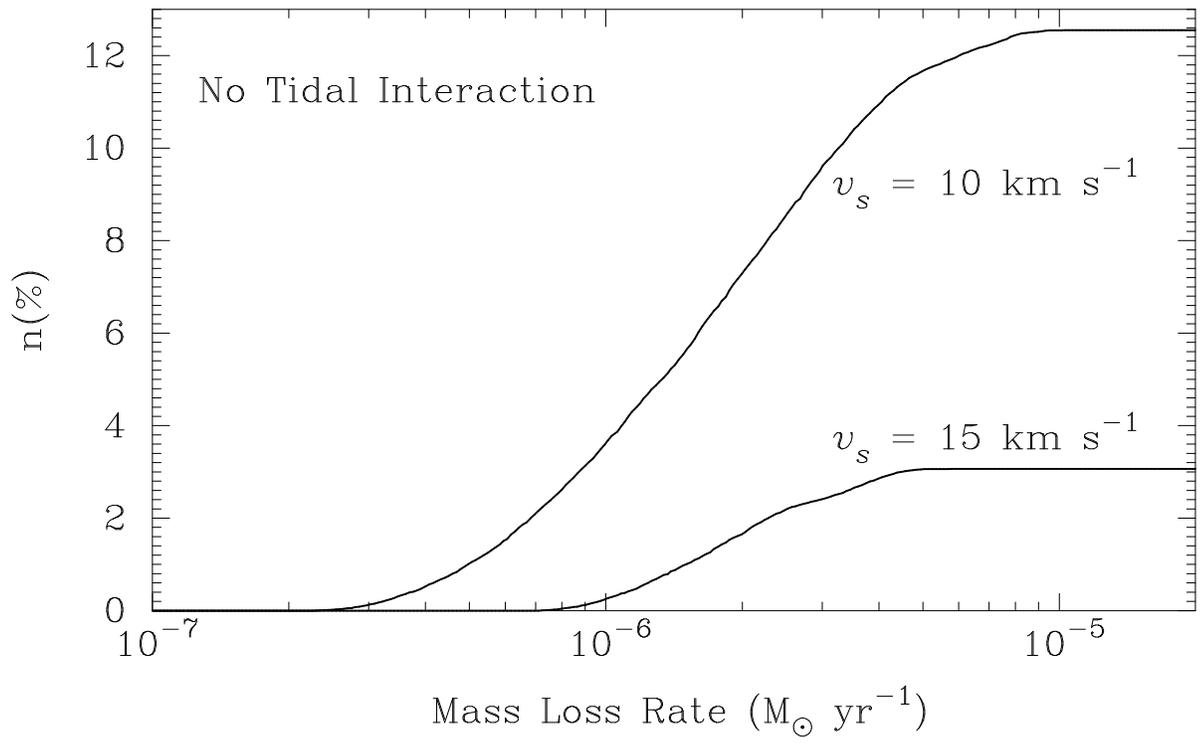
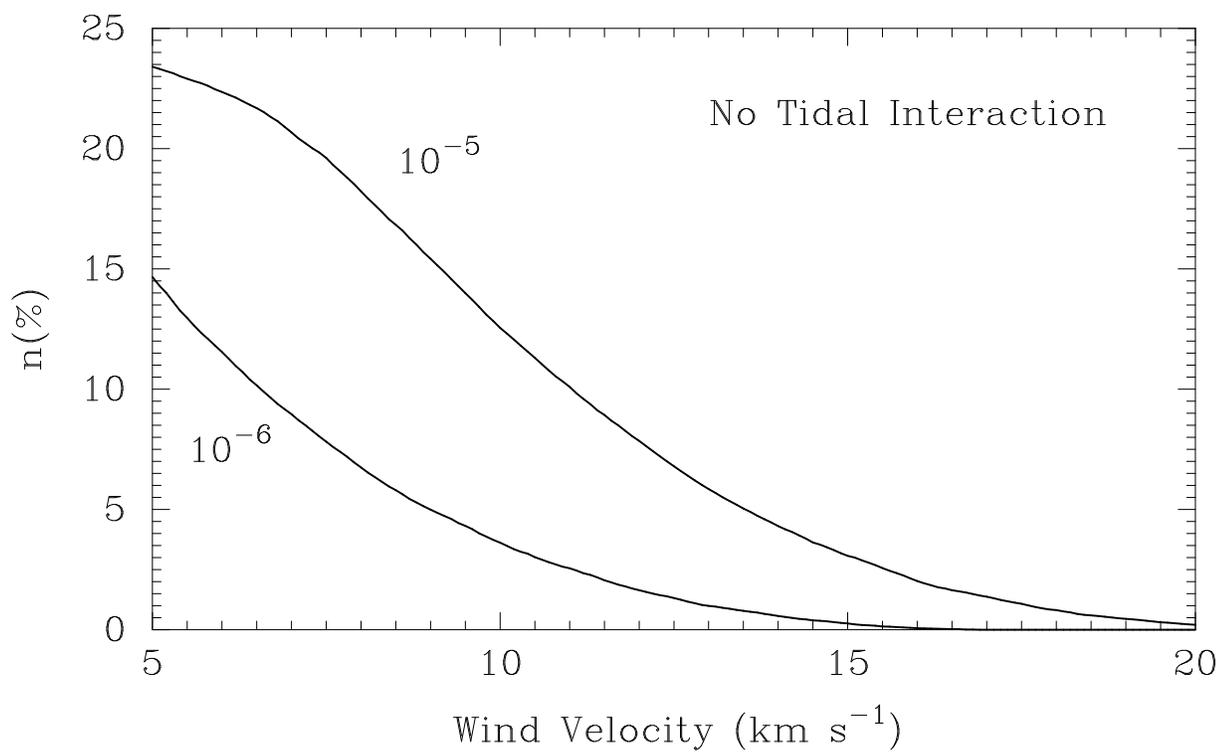
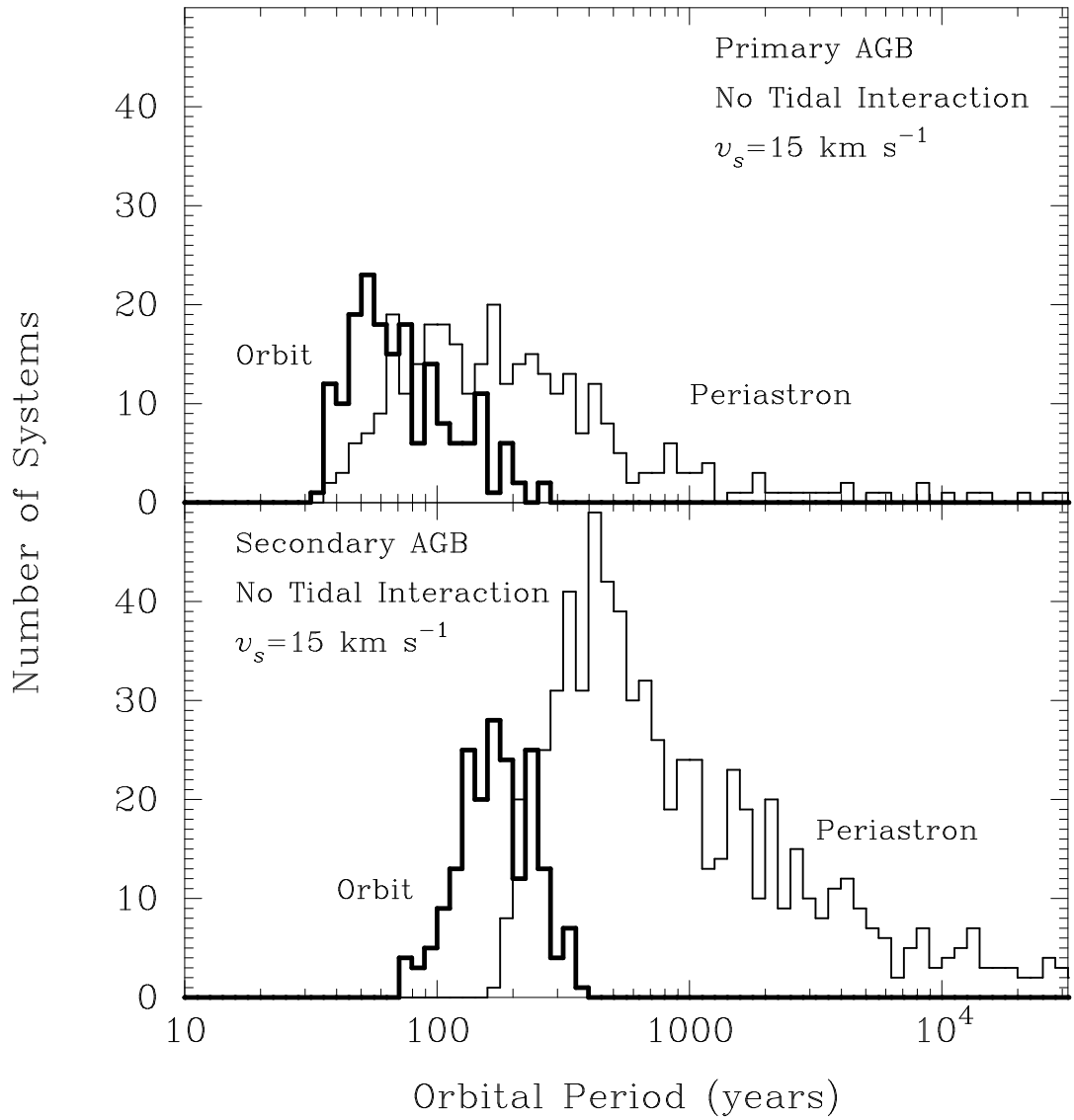
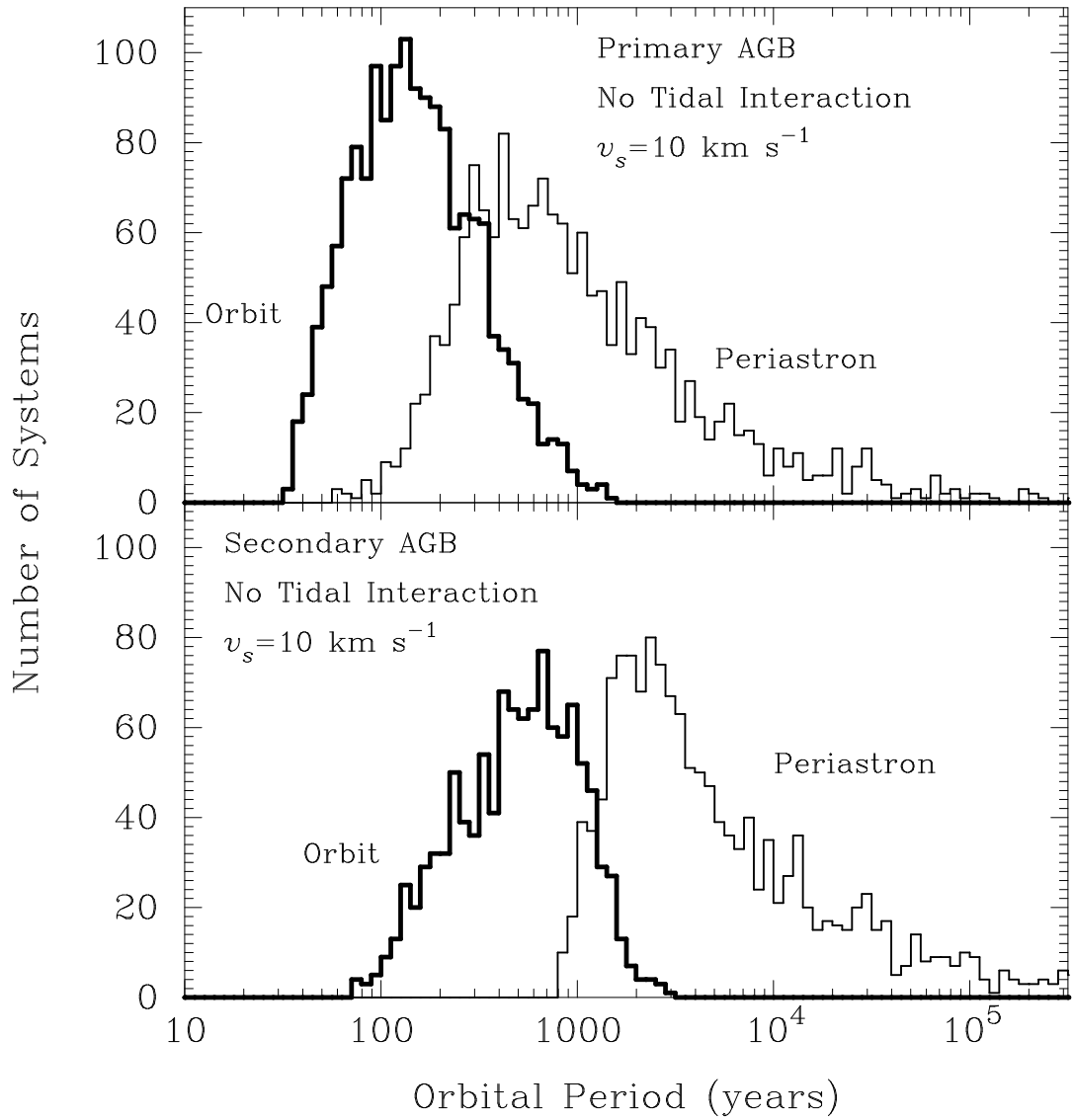
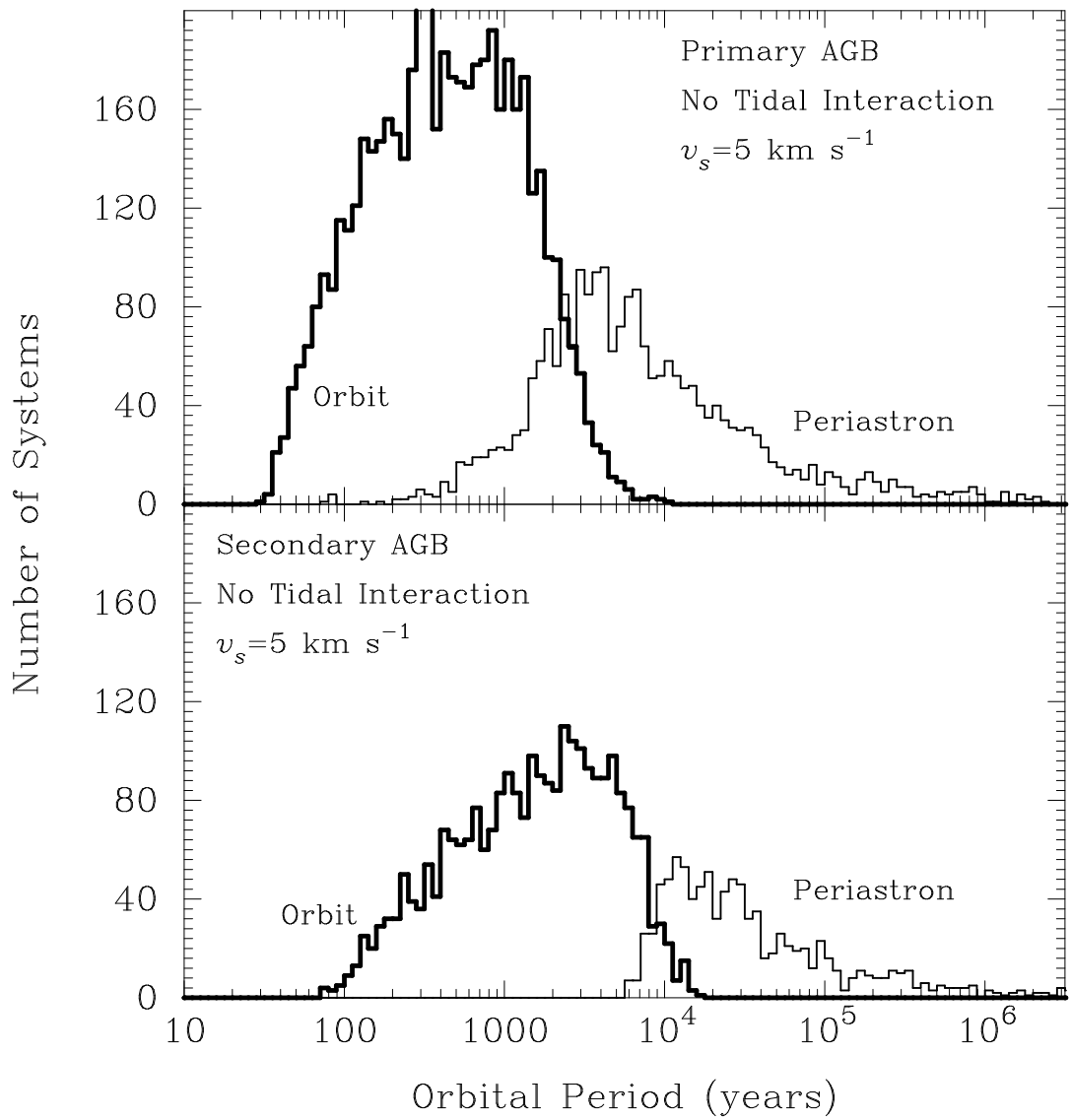
\end{document}